\pdfoutput=1

\documentclass[journal]{IEEEtran}
\usepackage{cite}

\usepackage[pdftex]{graphicx}
\graphicspath{{./graphics/}}
\DeclareGraphicsExtensions{.pdf,.png}

\usepackage[cmex10]{amsmath}
\usepackage[caption=false,font=footnotesize]{subfig}

\usepackage{fixltx2e}

\usepackage{url}

\begin{document}

\title{Temperature Effect and Correction Method of White Rabbit Timing Link}

\author{Hongming~Li,
        Guanghua~Gong,
        Weibin~Pan,
        Qiang~Du,
        Jianmin~Li
\thanks{Manuscript received May 22, 2014. This work was supported in part by the National Science Foundation of China (No. 11275111), the State Key Laboratory of Particle Detection and Electronics, Open Research Foundation of State Key Lab. of Digital Manufacturing Equipment \& Technology in Huazhong University of Science \& Technology.}
\thanks{Guanghua~Gong, is with the Key Laboratory of Particle \& Radiation Imaging (Tsinghua University), Ministry of Education, State Key Laboratory of Particle Detection and Electronics（Institute of High Energy Physics, CAS and University of Science and Technology of China） (telephone: 0086-10-62784552, e-mail: ggh@tsinghua.edu.cn).}
\thanks{Hongming~Li,Weibin~Pan, Jianmin~Li is with the Key Laboratory of Particle \& Radiation Imaging (Tsinghua University), Ministry of Education.(e-mail: lihm-13@mails.tsinghua.edu.cn)}
\thanks{Qiang Du is with Engineering Division, Lawrence Berkeley National Lab, QDu@lbl.gov)}
}

\markboth{IEEE TRANSACTION ON NUCLEAR SCIENCE}%
{Shell \MakeLowercase{\textit{et al.}}: Temperature Effect and Correction Method of White Rabbit Timing Link}


\maketitle

\begin{abstract}
To guarantee the angular resolution, the Large High Altitude Air Shower Observatory (LHAASO) requires a 500ps (rms) timing synchronization among the 6866 detect units for its KM2A sub-detector array. The White Rabbit technology is applied which combines sub-nanosecond precision timing transfer and gigabit Ethernet data transfer over the same fiber media. Deployed on a wild field at 4300m a.s.l. altitude, the WR network must maintain the precision over a wide temperature range. The temperature effect on a small WR link is measured, and contributions from different components like optical fiber, SFP module, fixed delay on PCB and ICs are separately studied and analyzed. An online real-time temperature correction method was applied based on the result which significantly reduce the synchronization variation from 300 ps to 50 ps in a temperature range of 50 degrees centigrade.
\end{abstract}



\section{Introduction}
\IEEEPARstart{S}{earching} the origin of galactic cosmic rays above 30TeV with high sensitivity and wide spectrum, the Large High Altitude Air Shower Observatory (LHAASO) project is a dedicated instrument consists of 4 sub-detector arrays. The square KM complex array includes 5635 scintillation electron and 1221 muon detectors covering the area of 1.2km$^2$. To guarantee the angular resolution of reconstructed air shower event, a 500ps (rms) timing synchronization must be achieved among the spreaded thousands of detectors \cite{cutewr:lhaaso}\cite{cutewr:duq}.

White Rabbit(WR) network, a fully deterministic Ethernet-based optical network capable of general purpose data transfer and synchronization, is applied for LHAASO KM2A timing network\cite{cutewr:wrproject}. WR technology is a combination of IEEE1588 (PTP) with two further enhancements: precise knowledge of the link delay and clock syntonization over the physical layer with Synchronous Ethernet(SyncE) \cite{cutewr:specification}. 

\subsection{synchronization methods}
\begin{figure}[!t]
\centering
\includegraphics[width=2.3in]{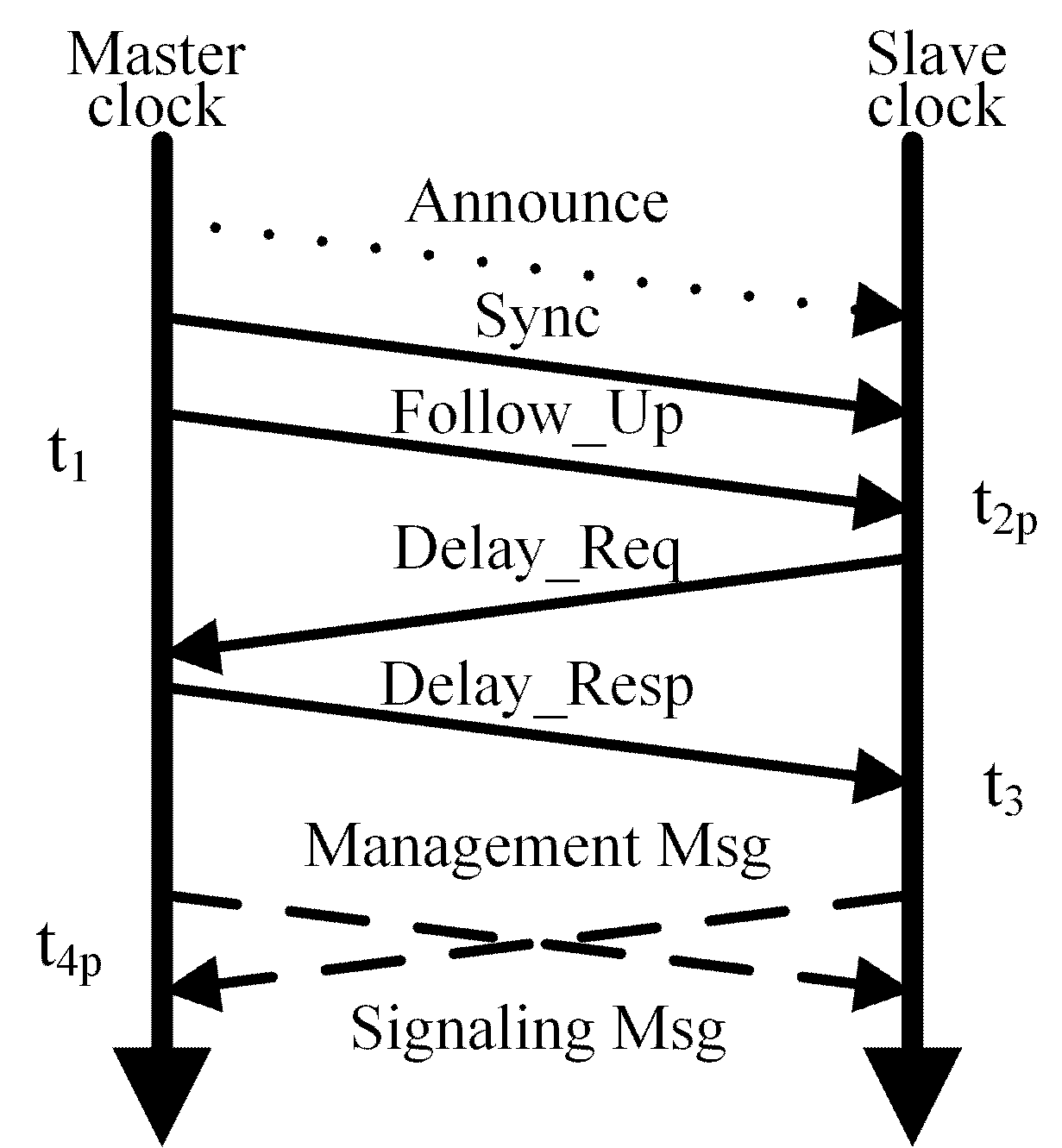}
\caption{PTP protocol used by WRPTP\cite{cutewr:tomas}}
\label{fig:ptp}
\end{figure}

\begin{figure}[!t]
\centering
\includegraphics[width=3.49in]{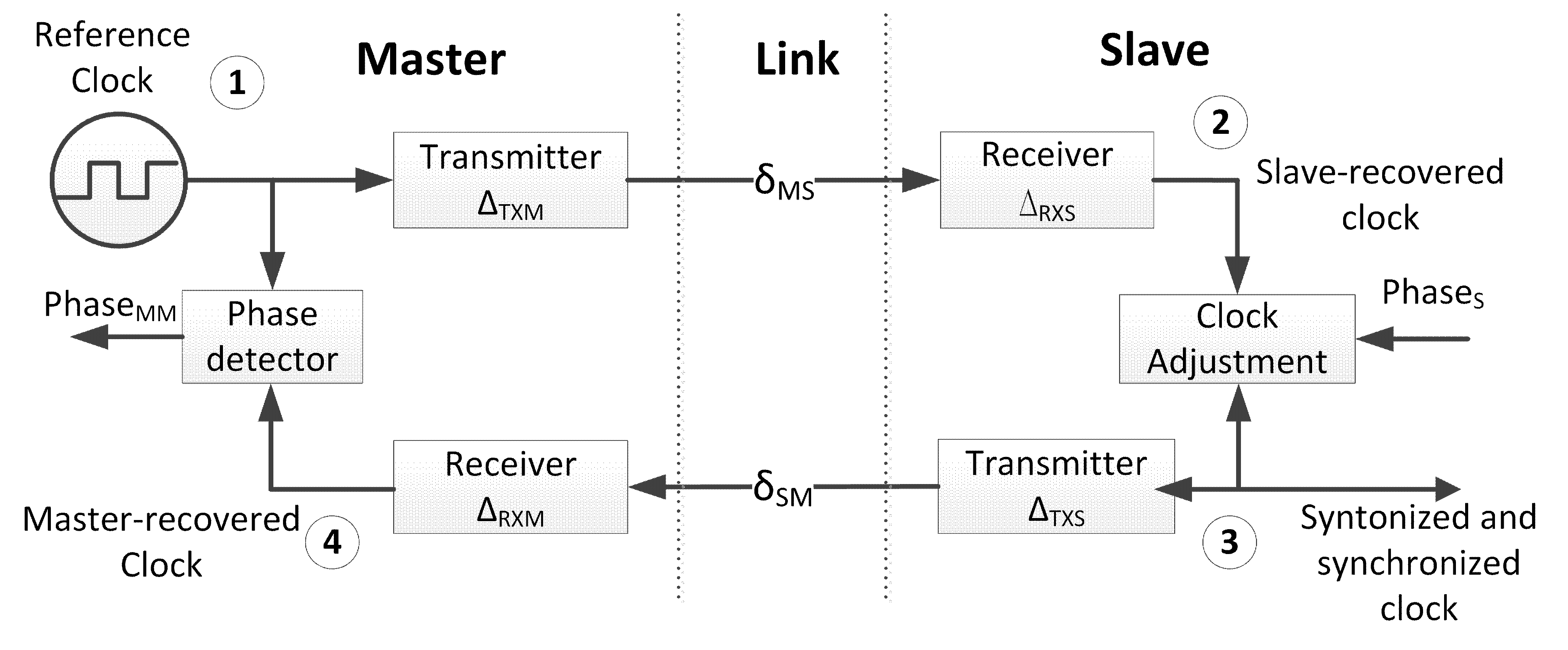}
\caption{WR link model \cite{cutewr:tomas}}
\label{fig:linkmodel}
\end{figure}

A WR link is formed by a pair of nodes, master and slave. Abosolute time synchronization is achieved by adjusting the clock phase and offset of the slave to that of the master, by applying an enhanced precision time protocol (PTP) which exchanges timestamp messages between master and slave as illustrated in Fig.\ref{fig:ptp}. Layer 1 clock syntonization(SyncE) and Digital Dual Mixer Time Difference (DDMTD) phase detector are used to improve the synchronization precision to sub-ns, as well as the knowledge of the link model of WR illustrated in Figure \ref{fig:linkmodel}, which takes into account the link media delay for uplink and downlink, the fixed delay caused by optical transmitter/receiver separatly\cite{cutewr:specification}\cite{cutewr:tomas}.

The one-way master to slave delay $delay_{ms}$,  slave to master delay  $delay_{sm}$ and round-trip delay $delay_{mm}$ can be expressed as the sum respectively:
\begin{align}
\label{eq:delay}
delay_{ms} &= \Delta_{txm} + \delta_{ms} + \Delta_{rxs} \nonumber\\
delay_{sm} &= \Delta_{rxm} + \delta_{sm} + \Delta_{txs} \\
\label{eq:delaymm}
delay_{mm} &= delay_{ms} + delay_{sm}  \nonumber\\
&=(t_{4p} - t_1) - (t_3 - t_{2p})
\end{align}
where $\Delta_{\{txm,rxs,txs,rxm\}}$ indicate the fixed delays due to master's or slave's transmission and reception circuitry. $\delta_{\{ms,sm\}}$ are the variable delays incurred in the transmission medium. $t_{\{1,2p,3,4p\}}$ are timestamps obtained from PTP as shown in Figure \ref{fig:ptp}.

WR uses a single-mode fiber as bi-directional communication medium. Different wavelengths are used for transmitting and receiving the data and the $\delta_{ms}$ and $\delta_{sm}$ have a relationship expressed as:
\begin{equation}
\label{eq:alpha}
\alpha = \frac{\delta_{ms}}{\delta_{sm}} - 1 = \frac{n_{\lambda_1}}{n_{\lambda_2}}-1
\end{equation}

where $\alpha$ is considered as the fiber asymmetry coefficient, which however is highly implementation-dependent and must be measured by experiment. $n_{\lambda_1}$ and $n_{\lambda_2}$ are refractive indexes of transmitting and receiving wavelength for the master. Normally the value of $\alpha$ is very small. For example the $\alpha$ value for the G652.D fiber is only around $10^{-4}$ when $\lambda_1$ is 1490nm and $\lambda_2$ is 1310nm.

Then the $delay_{ms}$ are calculated as:
\begin{equation}
\label{eq:delayms}
delay_{ms}=\frac{1+\alpha}{2+\alpha}(delay_{mm}-\Delta )+\Delta_{txm}+\Delta_{rxs}
\end{equation}
where$\Delta=\Delta_{txs}+\Delta_{rxs}+\Delta_{txm}+\Delta_{rxm}$\cite{cutewr:tomas}.

If all the fixed delays and $\alpha$ were constant and precisely measured, the calculated $delay_{MS}$ woule be accurate and the clock offset between master and slave would be zero. However, the fixed delay and $\alpha$ are affected by environmental factors that lead to the deviation of $delay_{ms}$. The clock offset between master and slave equals to the value of $delay_{ms}$ deviation. By measuring the pulse per second (PPS) timing signals generated by master and slave according to their internal clock, the offset can be monitored.

\section{Temperature effect analysis in WR link}
In current WR sepcification, the fixed delays and $\alpha$ are specially calibrated and stored as constant values to calculate the one-way delay. It has been proved that the worst case deviation of the clock offset between master and slave is still below 470 ps under a varying temperature range between 0 and 50 $^\circ$C \cite{cutewr:torturereport}. However, in the  situation of LASSHO, a 500ps (rms) timing synchronization among the 6866 detect units which are deployed on a wild field at 4300m a.s.l. altitude is required\cite{cutewr:lhaaso}. To maintain the precision with thousands of nodes and over a wide temperature range, more experiments and analysis should be done. In the report \cite{cutewr:torturereport} and our former tests, synchronization precision decreased with the varying temperature of the circuit boards and fibers, which can be essentially traced back to the variation of fixed delays and $\alpha$. 

\subsection{Variation of fixed delays}
The fixed delays contain propogation delays of PCB trace and electronic elements, optic/electric convert and FPGA logic delays. The temperature variation has a considerable effect on these factors, and leads to the fluctuation of the fixed delays.

If only the temperature of the slave nodes change, from Eq. \ref{eq:delayms}  we can get the variation of pps offset:
\begin{align}
\label{eq:ppsoffset}
offset=\delta_{delay_{ms}} &=\frac{1}{2+\alpha}\delta_{\Delta_{rxs}}-\frac{1+\alpha}{2+\alpha}\delta_{\Delta_{txs}} \nonumber\\
&\approx \frac{1}{2}(\delta_{\Delta_{rxs}} - \delta_{\Delta_{txs}}) 
\end{align}

where $offset$ denotes the PPS offset between master and slave, $\delta_{delay_{ms}}$ denotes the deviation of $delay_{ms}$, $\delta_{\Delta_{t(r)xs}}$ is the variation of slave’s transmitter (receiver) fixed delays. The approximation is due to the small value of $\alpha$.

From Eq. \ref{eq:ppsoffset}, any fluctuation of $\Delta_{t(r)xs}$ directly affect the offset adjustment hense the synchronization precision.

\subsection{Variation of alpha}
From the definition in Eq.\ref{eq:alpha}, $\alpha$ depends on the refractive indexes of certain wavelengths. There are two reasons that could cause the variation of $\alpha$.

First, the refractive index of certain wavelength changes with temperature. Since the WR specification applies wavelength division multiplexing for transmitting and receiving, the refractive index temperature dependency for different wavelengths are not identical thus could influence the value of $\alpha$ according to Eq.\ref{eq:alpha}. 

Second, the spectrum of the light generated by laser diode has temperature dependency too. The fluctuation of light wavelength can lead to the variation of $\delta_{ms}$ or $\delta_{sm}$, which causes the change of $\alpha$.

The temperature dependency of $\alpha$ coefficient factor can seriously degrade the synchronization precision in certain situation where the distance is long or temperature variation is large. The partial derivative of $delay_{ms}$ with respect to $\alpha$ is calculated as Eq.\ref{eq:delaymsPartial}.
\begin{equation}
\label{eq:delaymsPartial}
\frac{\partial (delay_{ms})}{\partial \alpha} = \frac{1}{(2+\alpha)^2}(delay_{mm}-\Delta_{txs}-\Delta_{rxm})
\end{equation}

From Eq.\ref{eq:delayms} and Eq.\ref{eq:delaymsPartial} we can get that: 
\begin{equation}
\label{eq:ppsoffsetR}
offset=\delta_{delay_{ms}}\approx\frac{1}{4}delay_{mm} \delta_{\alpha}
\end{equation}
where $\delta_{\alpha}$ is the variation of $\alpha$. The approximation is due to the trivial values of $\Delta_{txs}$ and $\Delta_{rxm}$ compared to $delay_{mm}$.

As the $delay_{mm}$ is mainly dominated by the length of the fiber connecting the master and slave, the longer the fiber is, the more significant the synchronization degrades for certain fluctuation of $\alpha$.

Several tests are designed to individually verify and further evaluate the influence of the temperature factors that affect the synchronization precision. The test setups and results are detailly described and discussed below.

\begin{figure*}[!t]
\subfloat[test1]{\includegraphics[width=1.8in]{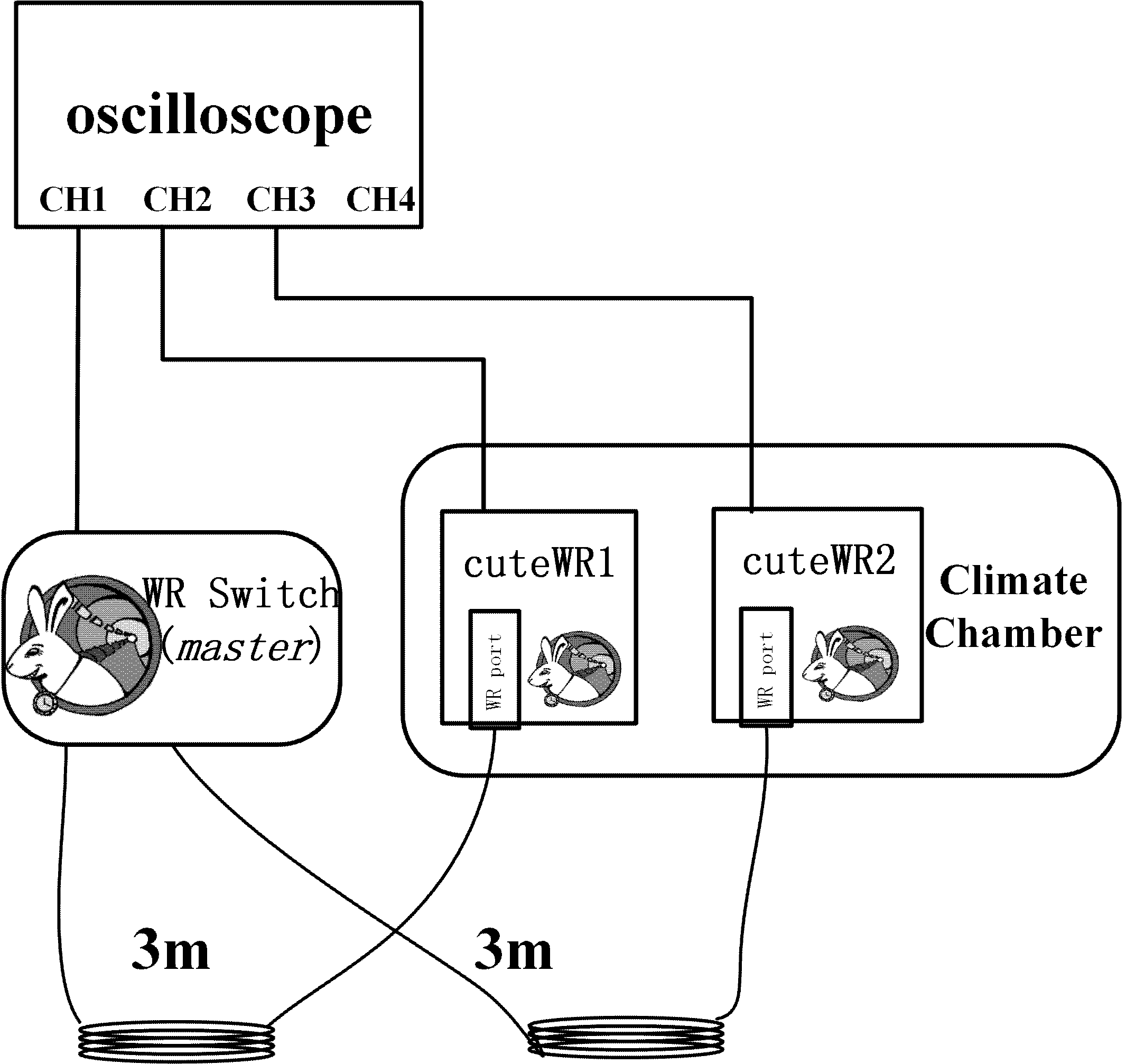}%
\label{fig:test1}}
\hfil
\subfloat[test2]{\includegraphics[width=1.8in]{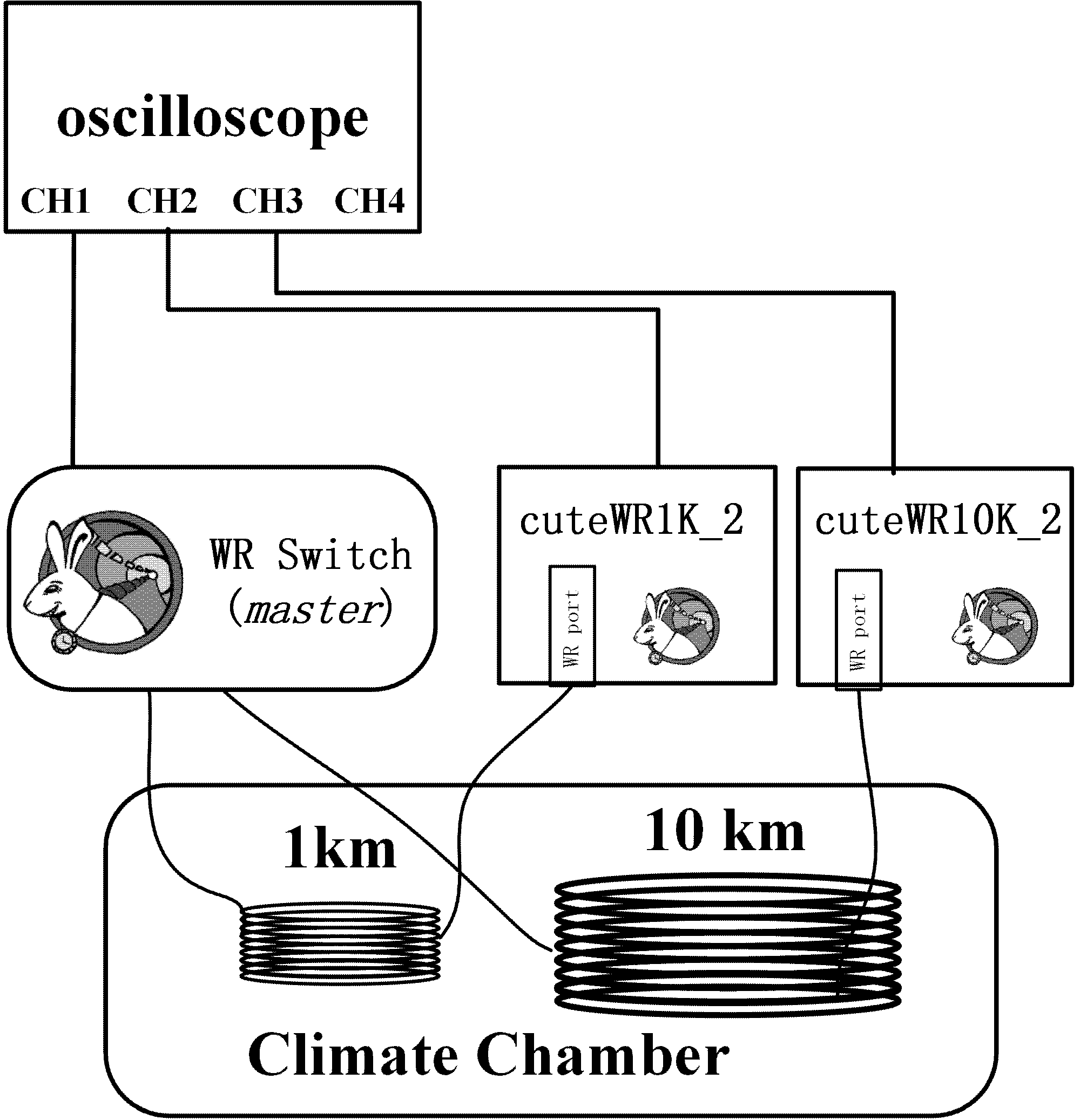}%
\label{fig:test2}}
\hfil
\subfloat[test3]{\includegraphics[width=1.8in]{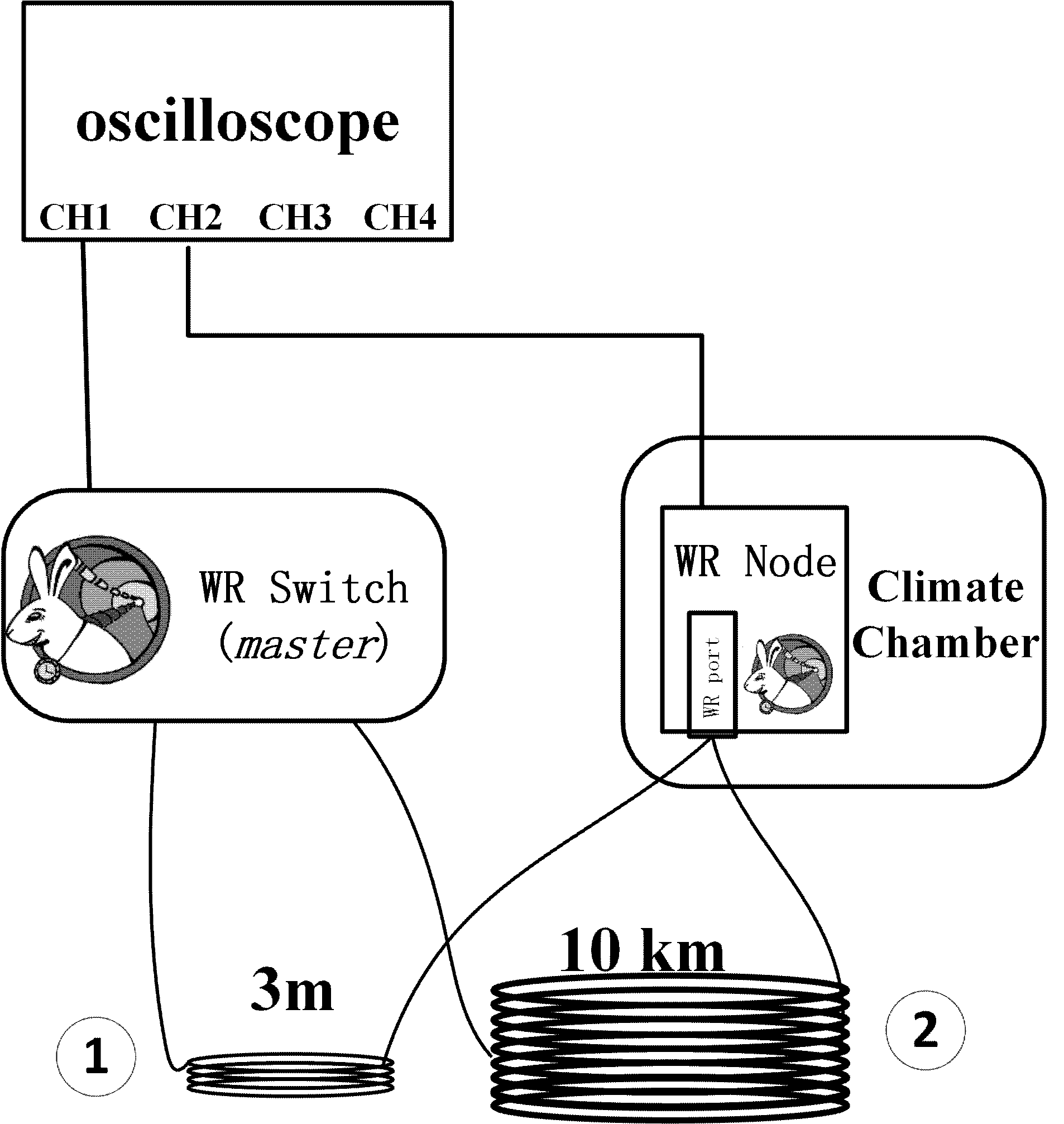}%
\label{fig:test3}}
\caption{Diagram of tests.}
\label{fig:testtopology}
\end{figure*}

\section{Temperature effect tests}
\subsection{System Under Test (SUT)}
A compact universal timing endpoint based on WR (CUTE-WR), is a simplified WR node designed as FPGA Mezzanine Card (FMC) with minimum components required \cite{cutewr:panwb}. In the tests, the CUTE-WR node acts as slave on the WR link, while the WR switch acts as master. As the CUTE-WR card and WR switch use the same control logic and auxiliary circuit for synchronization, the temperature feature of both should be similar, thus only the temperature effect on CUTE-WR card is studied in this paper. 

In the following tests, CUTE-WR cards or fibers are placed in the climate chamber, while the WR switch, oscilloscope and computer are placed in the stable laboratory conditions with ambient temperature of $25 \pm2^\circ$C. Fibers type G652.D and small form-factor pluggable transceiver(SFP) producted by Axcen are used, module type \& parameters and usage scenario are listed in Table \ref{tab:SFP1} - \ref{tab:SFP2}. The PPS offset values between the master switch and the CUTE-WR slave nodes are measured using a LeCroy oscilloscope with sampling rate of 10Gs/s. Other parameters and temperature on board are logged by PC via serial port. Detail deployment diagrams are shown in Figure \ref{fig:testtopology}.

\begin{table}[!t]
\renewcommand{\arraystretch}{1.3}
\caption{table of sfp information}
\label{tab:SFP1}
\centering
\begin{tabular}{|c|c|c|c|c|}
\hline 
 & modules type	   & Tx  	& Laser diode  & Rx   \\
\hline
SFP@1310  & AXGE-1254-0531 & 1310nm & FP  & 1490nm\\
\hline
SFP@1490  & AXGE-3454-0531 & 1490nm & DFB & 1310nm\\
\hline 
\end{tabular}
\end{table}

\begin{table}[!t]
\renewcommand{\arraystretch}{1.3}
\caption{table of sfp usage in each test}
\label{tab:SFP2}
\centering
\begin{tabular}{|c|c|c|}
\hline
 		 & Master Switch   & CUTE-WR node \\
\hline
test1    & SFP@1490		   & SFP@1310  \\
\hline
test2    & SFP@1490 	   & SFP@1310  \\
\hline
test3.1  & SFP@1490		   & SFP@1310 \\
\hline
test3.2	 & SFP@1310		   & SFP@1490 \\
\hline
\end{tabular}
\end{table}

Test 1, shown in Figure \ref{fig:test1}, includes two CUTE-WR cards (denoted as cuteWR1 and cuteWR2) connected independently to the WR switch's different ports. CUTE-WR cards are placed in climate chamber to test the influence of the fixed delay variation. Fibers of 3 meters are used to minimize the effect of the $\alpha$ variation.

Test 2, shown in Figure \ref{fig:test2}, includes two CUTE-WR cards connected independently to the WR switch's different ports with fibers of 1km and 10km respectively  (denoted as cuteWR1k\_2 and cuteWR10k\_2). Fibers are placed in the climate chamber to test the influence of the $\alpha$ variation.

Test 3, shown in Figure \ref{fig:test3}, the single CUTE-WR card is connected to WR switch with different length of fibers which are 3m (denoted as CuteWR3\_3) and 10km (denoted as CuteWR10k\_3). To test the influence of the $\alpha$ variation caused by the light spectrum temperature dependency of laser diode, the CUTE-WR card is placed in climate chamber and the fixed delays variation has been corrected with methods discussed in section III.B. The test is repeated after swapping the SFP modules of the master switch and slave nodes to evaluate the behavior of different laser diode types.

\subsection{Temperature effect of fixed delay and correction method}
\begin{figure}[!t]
\centering
\includegraphics[width=2.5in]{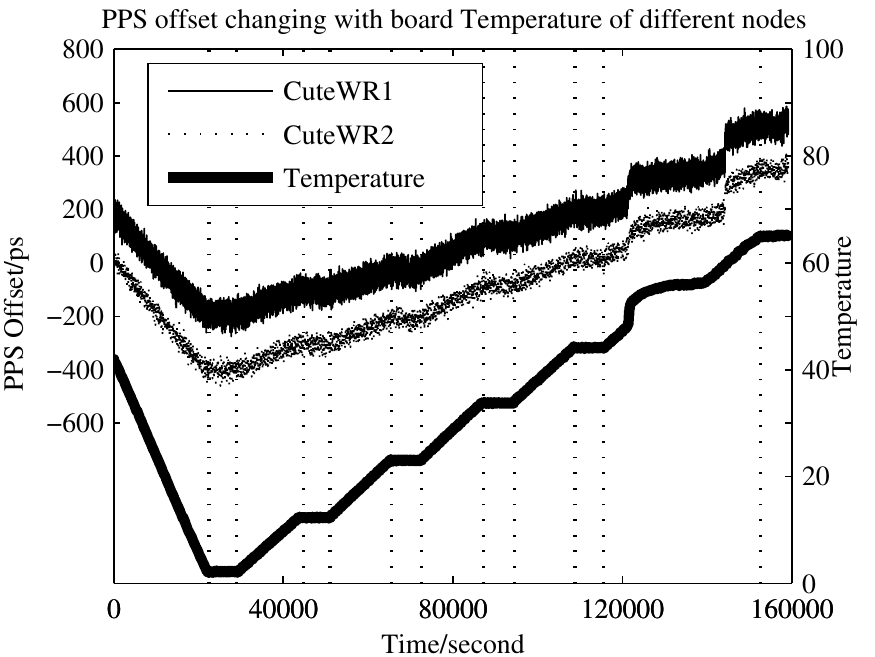}
\caption{PPS offset between the master and the CUTE-WR slave nodes changing over time when CUTE-WR board temperature varying (result of test1)}
\label{fig:offset_brd_temp}
\end{figure}

Figure \ref{fig:offset_brd_temp} shows the results of Test1. The degradation of synchronization performance (PPS offset) can be about 700 ps when CUTE-WR cards are under a varying ambient temperature between -10 and 55 $^\circ$C.

\begin{figure}[!t]
\centering
\includegraphics[width=3.0in]{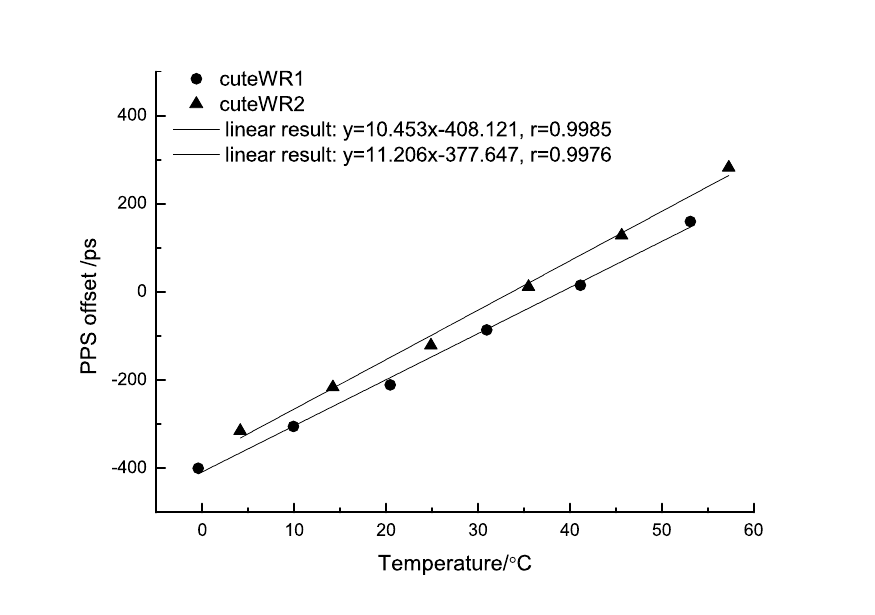}
\caption{PPS offset between the master and the CUTE-WR slave nodes with temperatures}
\label{fig:ppsoffset}
\end{figure}
\begin{figure}[!t]
\centering
\includegraphics[width=3.0in]{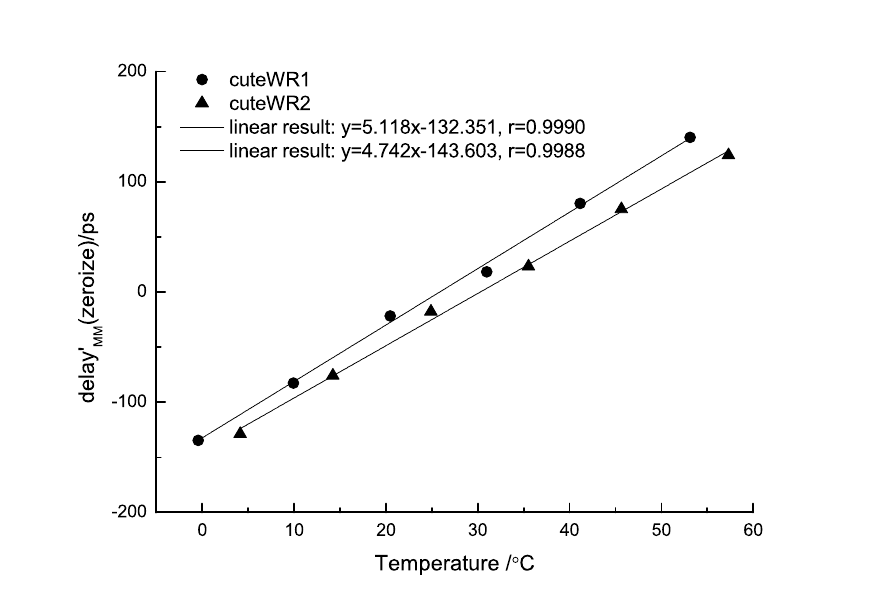}
\caption{Round-trip delay between the master and the CUTE-WR slave nodes with temperatures}
\label{fig:delaymm}
\end{figure}

There is an approximate linear correlation between the PPS offset and temperature as shown in Figure \ref{fig:ppsoffset}. As shown in Figure \ref{fig:delaymm}, the round-trip delay $delay_{mm}$ shows a linear tempearture dependency too.

As only the CUTE-WR nodes are placed in the climate chamber, from Eq.\ref{eq:ppsoffset} we infer that the coefficient of PPS offset changes with temperature $\tau_{offset}$ is:
\begin{equation}
\label{eq:ppsoffsetS}
\tau_{offset} = \frac{1}{2}(\tau_{\Delta_{rxs}} - \tau_{\Delta_{txs}})
\end{equation}

where $\tau_{\Delta_{txs}}$ and $\tau_{\Delta_{rxs}}$ are the temperature coefficients of the fixed delays of slave transmit and receive, which have linear temperature depedency under the first order approximation.

From Eq. \ref{eq:delay} and Eq. \ref{eq:delaymm} the variation of $delay_{mm}$ in test1 is merely caused by the fixed delays of slave nodes. The coefficient of $delay_{mm}$ changes with temperature $\tau_{delay_{mm}}$ is:
\begin{equation}
\label{eq:delaymm1}
\tau_{delay_{mm}} = \tau_{\Delta_{txs}} + \tau_{\Delta_{rxs}}
\end{equation}  

The $delay_{mm}$ is measured by master and can be readout via serial port while the offset can be measured by oscilloscope. By collection of multi data sets for $delay_{mm}$ and offset under different temperatures, the value of $(\tau_{\Delta_{rxs}}-\tau_{\Delta_{txs}})$ and $(\tau_{\Delta_{txs}}+\tau_{\Delta_{rxs}})$ can be easily obtained by linear fitting, thus we can get the value of $\tau_{\Delta_{txs}}$ and $\tau_{\Delta_{rxs}}$.

From Figure \ref{fig:ppsoffset}$\sim$\ref{fig:delaymm}, we can get the following results:
\begin{align}
\label{eq:deltaValue}
\tau_{\Delta_{txs1}} & =-7.9 \,\mathrm{ps/^{\circ}C},\qquad \tau_{\Delta_{rxs1}}= 13.0 \,\mathrm{ps/^{\circ}C} \nonumber\\ 
\tau_{\Delta_{txs2}} & =-8.8 \,\mathrm{ps/^{\circ}C},\qquad \tau_{\Delta_{rxs1}}= 13.6 \,\mathrm{ps/^{\circ}C}
\end{align}

Taking the average value of $\tau_{\Delta_{rxs}}=13.3$ ps$/^{\circ}$C and $\tau_{\Delta_{txs}}=-8.4$ ps$/^{\circ}$C as the temperature coefficients for fixed delays of transmit and receive. We can adapt the fixed delay values as functional variable of the temperature measured by the thermometer on board to achieve the real-time temperature correction. The performance test will be discussed in section III.D.

\subsection{Temperature effect of alpha}
\subsubsection{Fiber link temperature effect}
\begin{figure}[!t]
\centering
\includegraphics[width=2.5in]{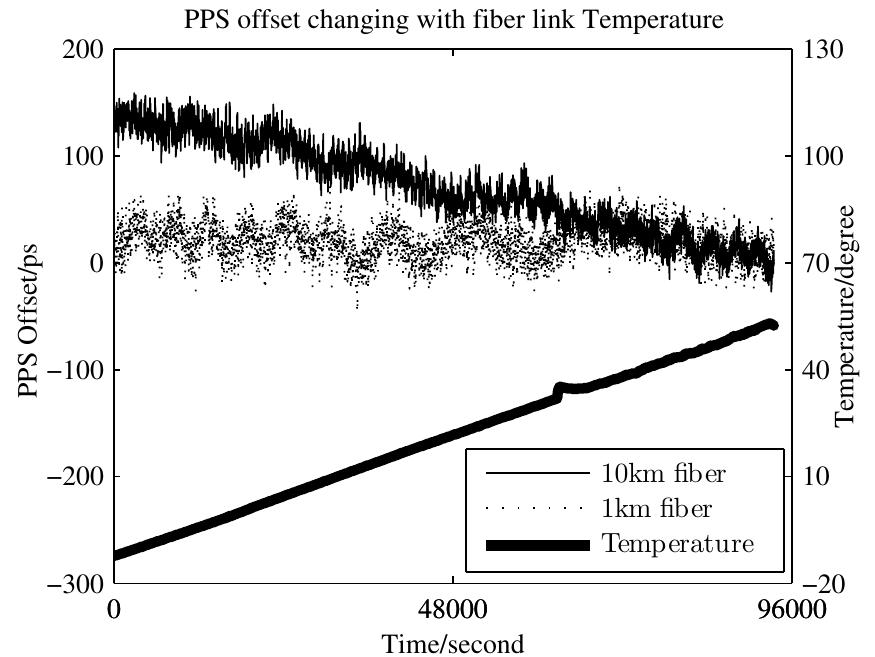}
\caption{PPS offset between the master and the CUTE-WR slave nodes changing over time when the fiber link temperature varying (result of test2)}
\label{fig:offset_link_temp}
\end{figure}

\begin{figure}[!t]
\centering
\includegraphics[width=2.5in]{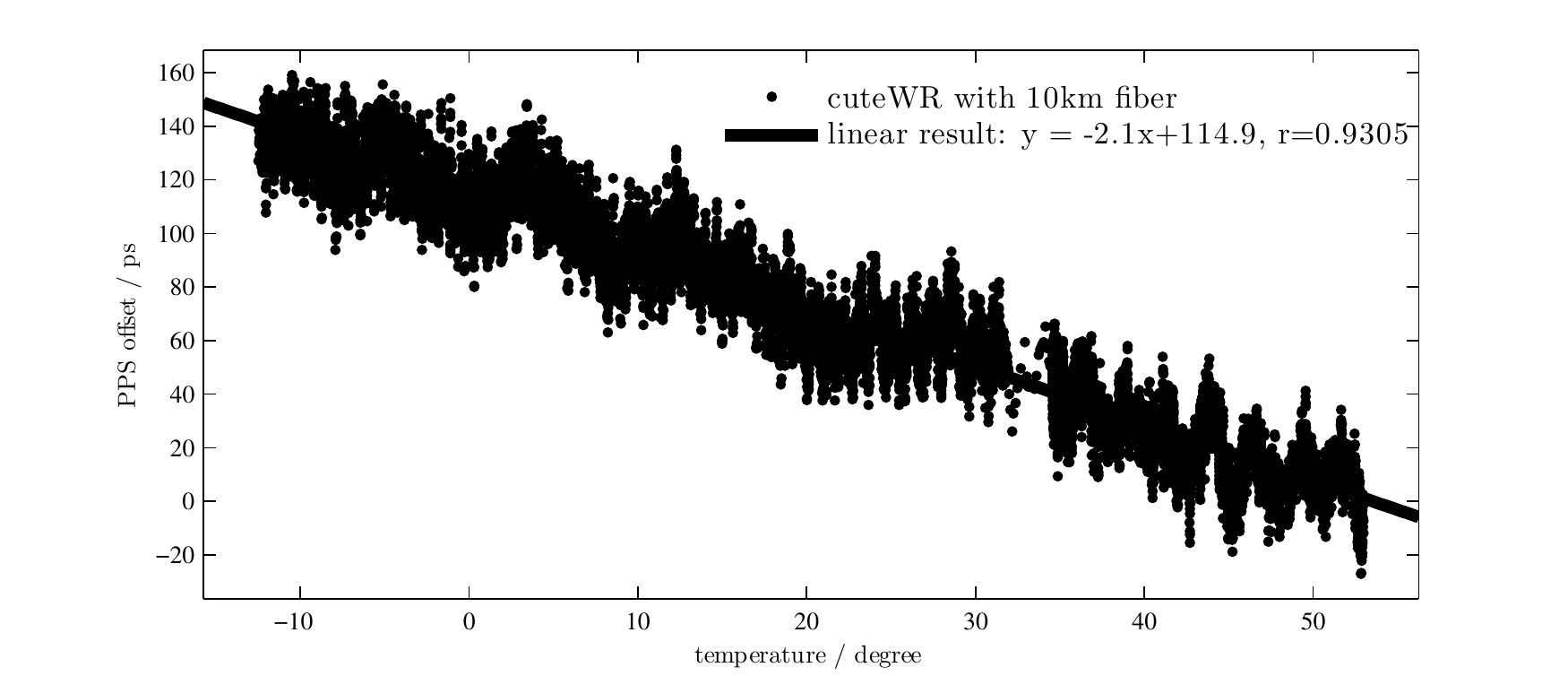}
\caption{PPS offset between the master and the CUTE-WR slave nodes changes with temperature for 10km fiber}
\label{fig:offset_link_temp_linear}
\end{figure}

Figure \ref{fig:offset_link_temp} shows the results of Test2. The PPS offset between the master switch and cuteWR1k\_2 shows no correlation with fiber link temperature, while that between the master switch and cuteWR10k\_2 has a linear correlation with fiber link temperature as shown in Figure \ref{fig:offset_link_temp_linear}, the coefficient is about $-$2.1 ps/$^{\circ}$C with a linear fit (R-square is 0.9305).

We can infer the temperature coefficient of PPS offset for fibers we used (G652.D) equals to $-$0.2 ps/($^{\circ}$C$\cdot$km). Then the calculated PPS offset of cuteWR1k\_2 changes 13 ps under a varying ambient temperature between -10 and 55 $^{\circ}$C, which is too small to be detected in the test.

\subsubsection{Optical transceiver temperature}
\begin{figure}[!t]
\centering
\subfloat[Fibers of 3m]{\includegraphics[width=1.7in]{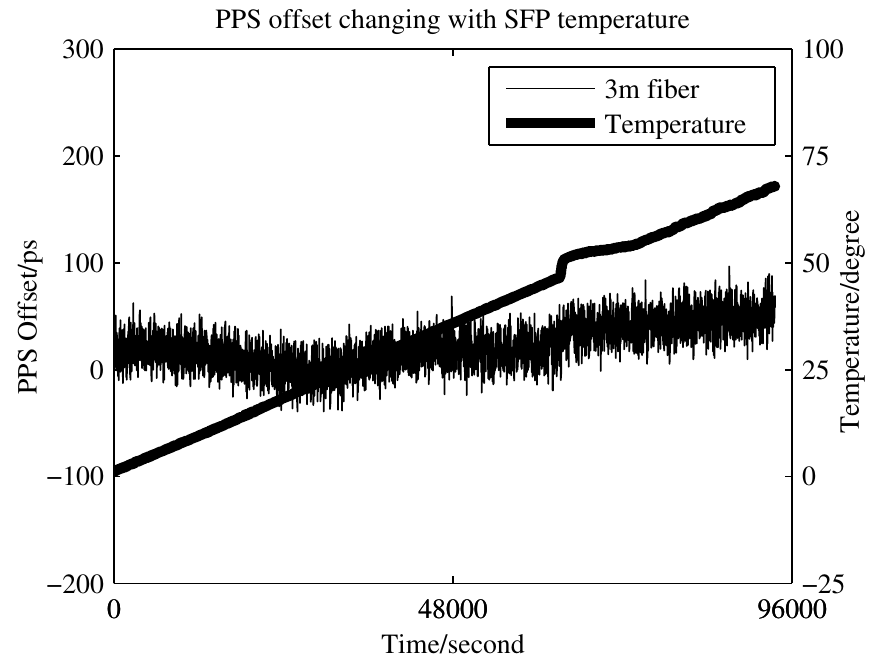}%
\label{fig:SFP1_3}}
\hfill
\subfloat[Fibers of 10km]{\includegraphics[width=1.7in]{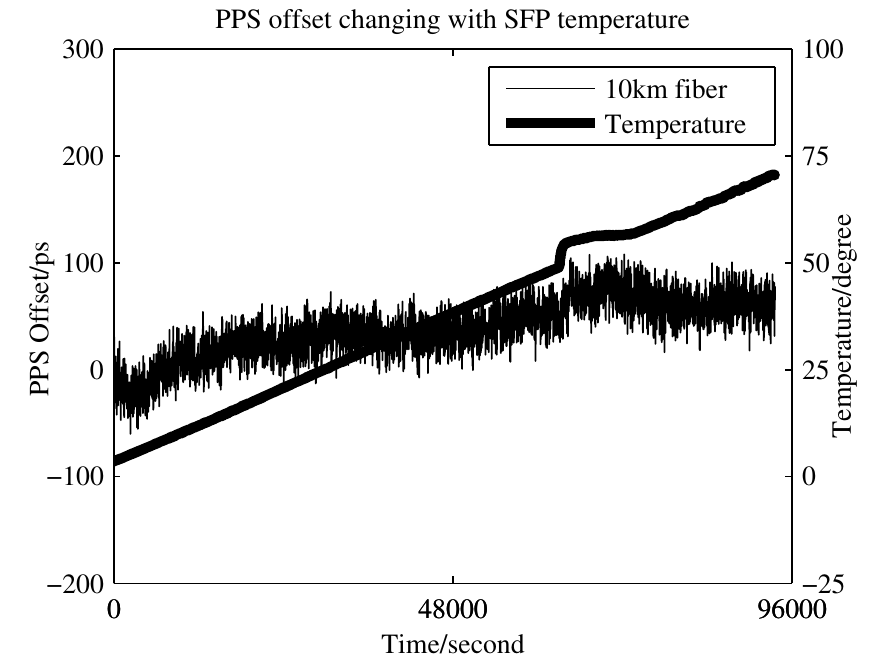}%
\label{fig:SFP1_10k}}
\caption{PPS offset between the master and the CUTE-WR slave nodes changing over time when the temperature of  SFP@1310 used in CUTE-WR card varying (results of test3.1)}
\label{fig:SFP1}
\end{figure}

\begin{figure}[!t]
\centering
\subfloat[Fibers of 3m]{\includegraphics[width=1.7in]{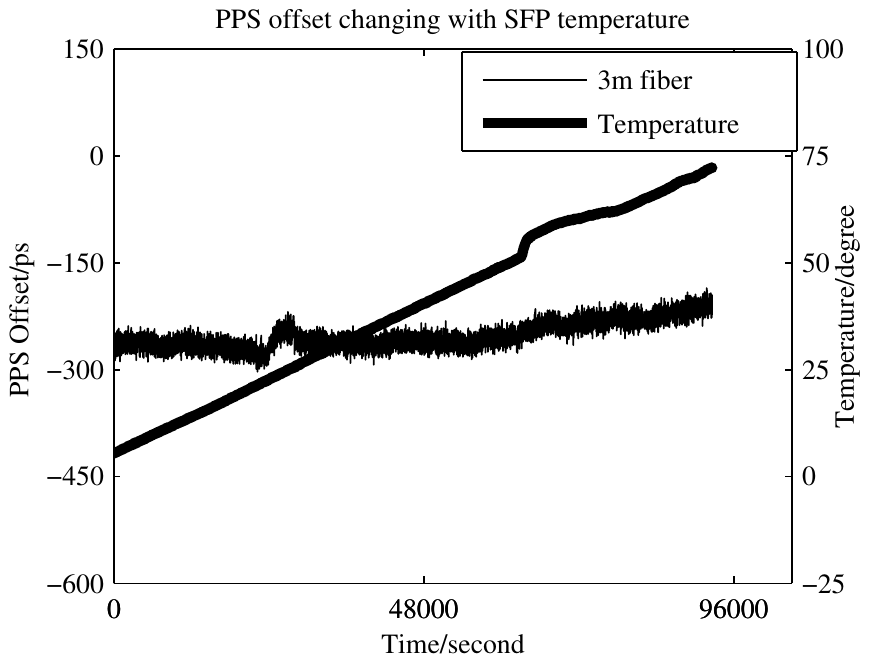}%
\label{fig:SFP2_3}}
\hfill
\subfloat[Fibers of 10km]{\includegraphics[width=1.7in]{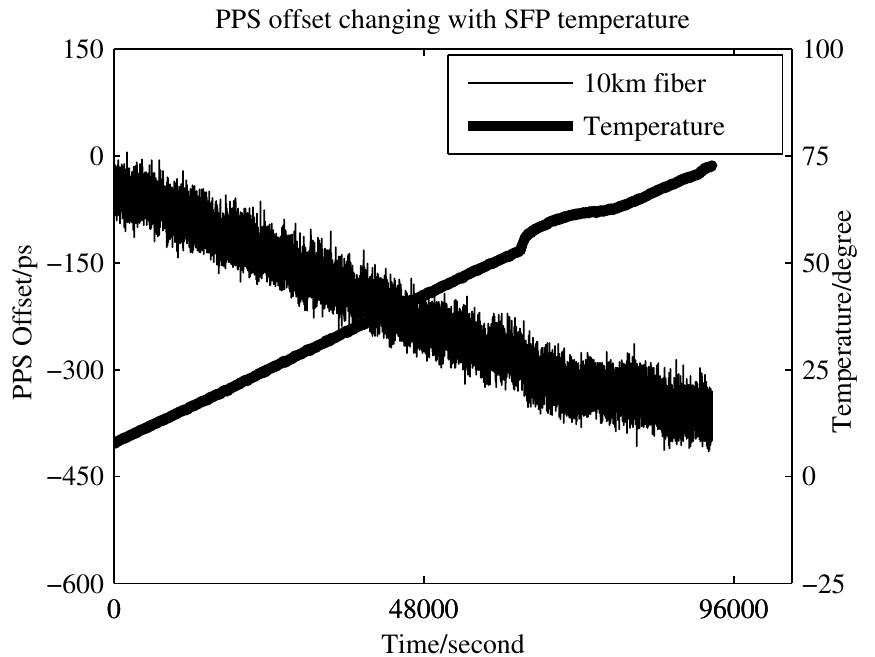}%
\label{fig:SFP2_10k}}
\caption{PPS offset between the master and the CUTE-WR slave nodes changing over time when the temperature of  SFP@1490 used in CUTE-WR card varying (results of test3.2)}
\label{fig:SFP2}
\end{figure}

Figure \ref{fig:SFP1}-\ref{fig:SFP2} show the results of Test3. In Figure \ref{fig:SFP1}, the SFP transceiver used by CUTE-WR node is SFP@1310 (test3.1), while in Figure \ref{fig:SFP2} SFP@1490 is used by CUTE-WR node (test3.2). The node's fixed delay has been corrected using the methods discussed in section III.B, which can be seen from Fig.\ref{fig:SFP1_3} and Fig.\ref{fig:SFP2_3}, that with 3m fiber, the PPS offset is not related with temperature. 

\begin{figure}[!t]
\centering
\subfloat[SFP@1310 used]{\includegraphics[width=1.7in]{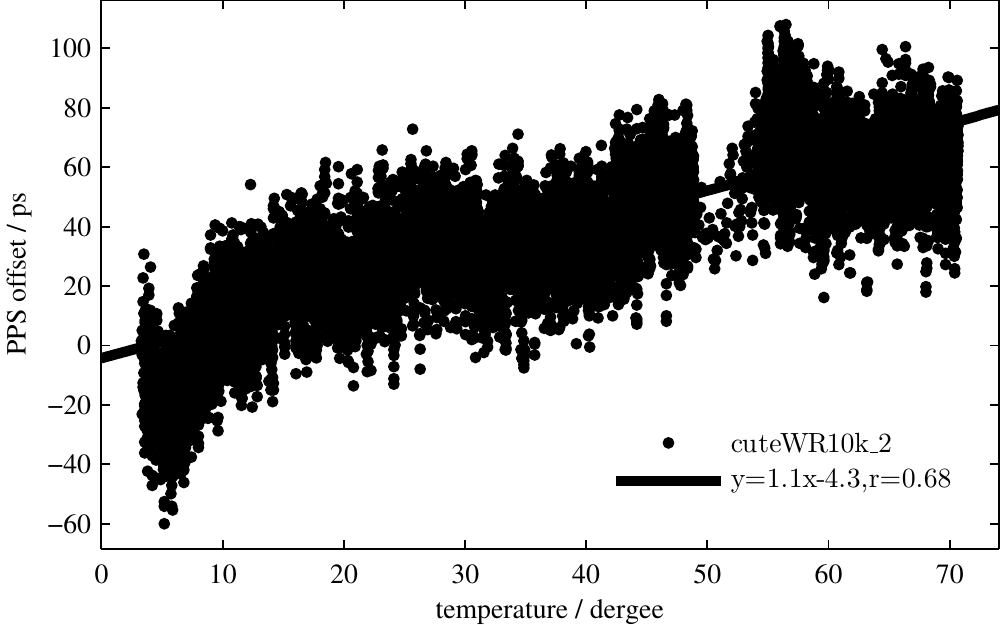}%
\label{fig:offset_link_sfp_linear1}}
\hfill
\subfloat[SFP@1490 used]{\includegraphics[width=1.7in]{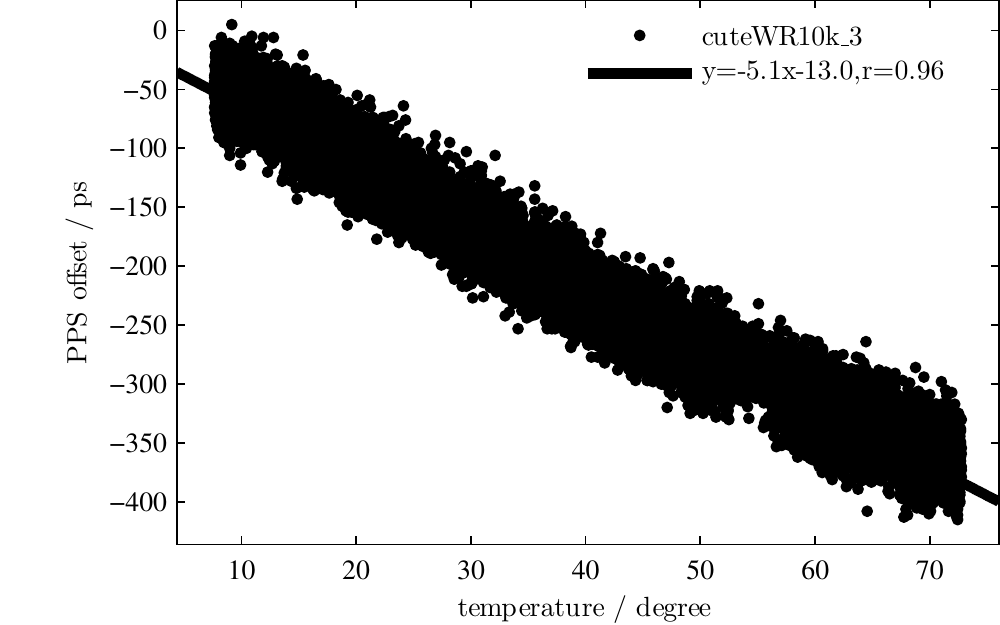}%
\label{fig:offset_link_sfp_linear2}}
\caption{PPS offset between the master and the CUTE-WR slave nodes changing with the SFP temperature used in the slave CUTE-WR card}
\label{fig:offset_link_sfp_linear}
\end{figure}

With 10 km fiber, the correlation is almost linear as shown in Fig \ref{fig:offset_link_sfp_linear}. The temperature coeffiecnt of PPS offset for cuteWR10k\_2 (SFP@1310) equals to 0.11 ps/($^{\circ}$C$\cdot$km) with a linear fit (R-square is 0.68) and that for cuteWR10k\_3 (SFP@1490) equals to $-$0.51 ps/($^{\circ}$C$\cdot$km) (R-square is 0.96). As the impact of the fixed delays variation is independent of fiber length, the PPS offset variation is only caused by $\alpha$. 

The results and coefficient values can be explained by the spectrum temperature dependency of the specific laser diode. SFP@1310 is a "Fabry-Perot" transmitter, whose temperature coefficient of the central wavelength is 0.4$\sim$0.5 nm/$^{\circ}$C \cite{cutewr:sfp1datasheet}. The SFP@1490 is a "Distributed Feedback" transmitter, whose temperature coefficient of the central wavelength is 0.08$\sim$0.12 nm/$^{\circ}$C \cite{cutewr:sfp2datasheet}. The chromatic dispersion of fiber G652.D can be calculated through the Eq.\ref{eq:ChromaticDispersion} provided by ITU-T G.652\cite{cutewr:itut}. The chromatic dispersion are $-$1.3$\sim$0.9 ps/(nm$\cdot$km) when $\lambda$ is 1310nm and 12.9$\sim$14.4 ps/(nm$\cdot$km) when $\lambda$ is 1490 nm. Then the temperature coefficients of the transmit delay $\tau_{\delta_{nm}}/\tau_{\delta_{sm}}$ are $-$0.65$\sim$0.45 ps/($^{\circ}$C$\cdot$km) for SFP@1310 and 1.03$\sim$1.73 ps/($^{\circ}$C$\cdot$km) for SFP@1490. 
\begin{equation}
\label{eq:ChromaticDispersion}
\frac{0.092\lambda}{4}[1-(\frac{1324}{\lambda})^4] \leq D(\lambda) \leq \frac{0.092\lambda}{4}[1-(\frac{1300}{\lambda})^4]
\end{equation}
where $D(\lambda)$ is the chromatic dispersion of wavelength $\lambda$, the unit of $\lambda$ is nm.

\begin{equation}
\label{eq:TxToPPS}
\begin{split}
&\tau_{offset} = \frac{1}{2}\tau_{\delta_{ms}}\\
&\tau_{offset} = -\frac{1}{2}\tau_{\delta_{sm}}
\end{split}
\end{equation}

The coeffients of PPS offset related to the SFP transmitter temperature are calculated as Eq. \ref{eq:TxToPPS}, which are $-$0.23$\sim$ 0.33 ps/($^{\circ}$C$\cdot$km) for SFP@1310 and $-$0.87$\sim-$0.52 ps/($^{\circ}$C$\cdot$km) for SFP@1490. 

The measured degradation of synchronization matches the calculation as listed in Table \ref{tab:SFPcoefficent}.
 
\begin{table}[!t]
\renewcommand{\arraystretch}{1.3}
\caption{table of temperature coefficients of fibers \& sfps}
\label{tab:SFPcoefficent}
\centering
\begin{tabular}{|c|c|c|}
\hline 
&\multicolumn{2}{c|}{Temp. Coefficient / ps/($^{\circ}$C$\cdot$km)}\\
\cline{2-3}
& Calculated & Measured   \\
\hline
G652.D(M-S1490nm,S-M1310nm) & - &-0.2\\
\hline 
SFP@1310(on slave)  & -0.23 $\sim$ 0.33& 0.11 \\
\hline
SFP@1490(on slave)  & -0.87 $\sim$ -0.52 & -0.51 \\
\hline
\end{tabular}
\end{table}

\subsection{Performance test}
To examine the synchronization perforamce of the fixed delays correction, two CUTE-WR cards (cuteWR1, cuteWR2) and another new one (donted as cuteWR3) are placed in climated chamber and connect to the master switch with fibers of 2km respectively. SFP@1310 is used in CUTE-WR card and SFP@1490 used for the switch. Fibers and the master switch are under room temperature. 

\begin{figure}[!t]
\centering
\subfloat[PPS offset of CuteWR1]{\includegraphics[width=1.6in]{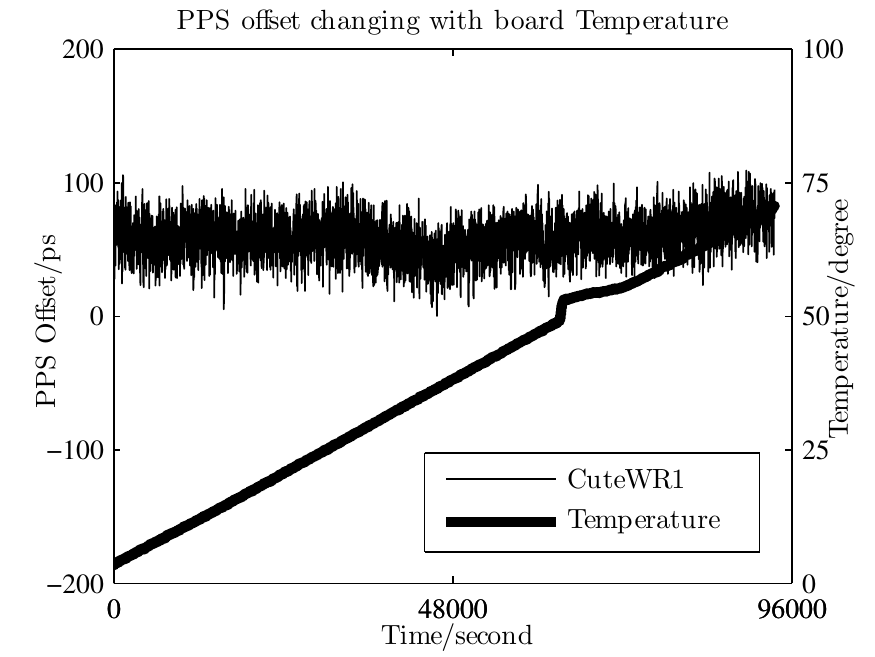}%
\label{fig:performance1}}
\subfloat[Histogram of CuteWR1]{\includegraphics[width=1.6in]{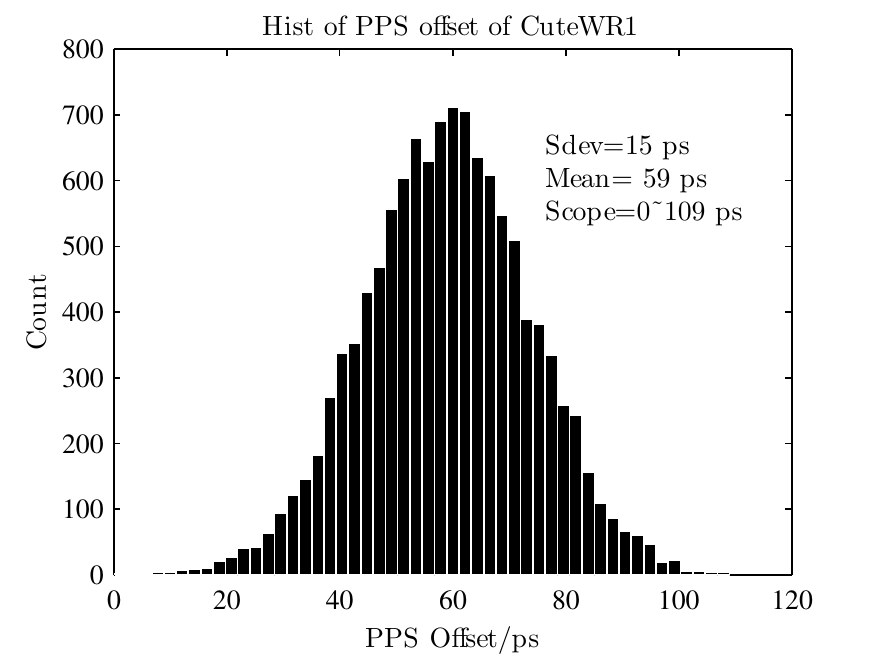}%
\label{fig:hist1}}
\hfill
\subfloat[PPS offset of CuteWR2]{\includegraphics[width=1.6in]{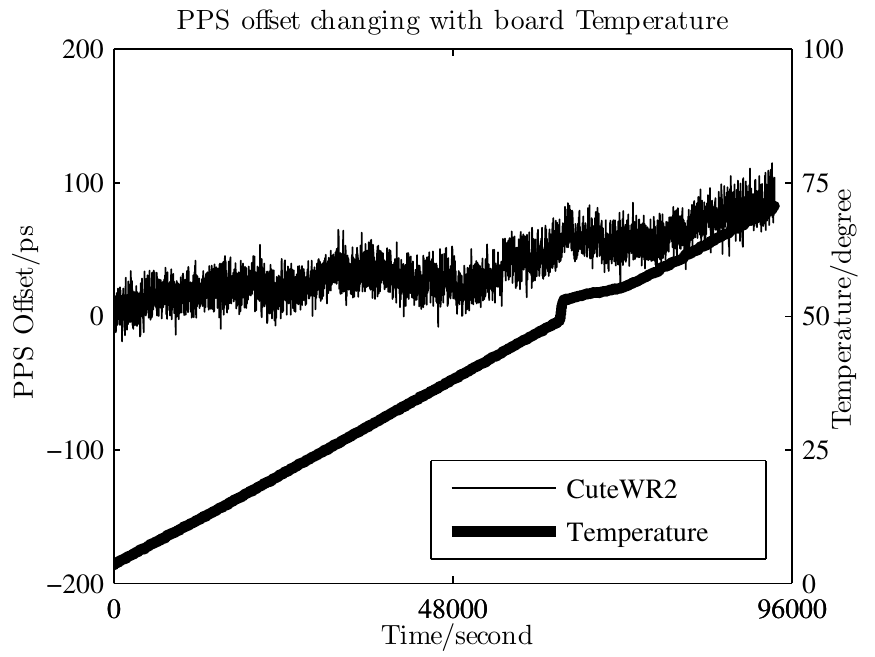}%
\label{fig:performance2}}
\subfloat[Histogram of CuteWR2]{\includegraphics[width=1.6in]{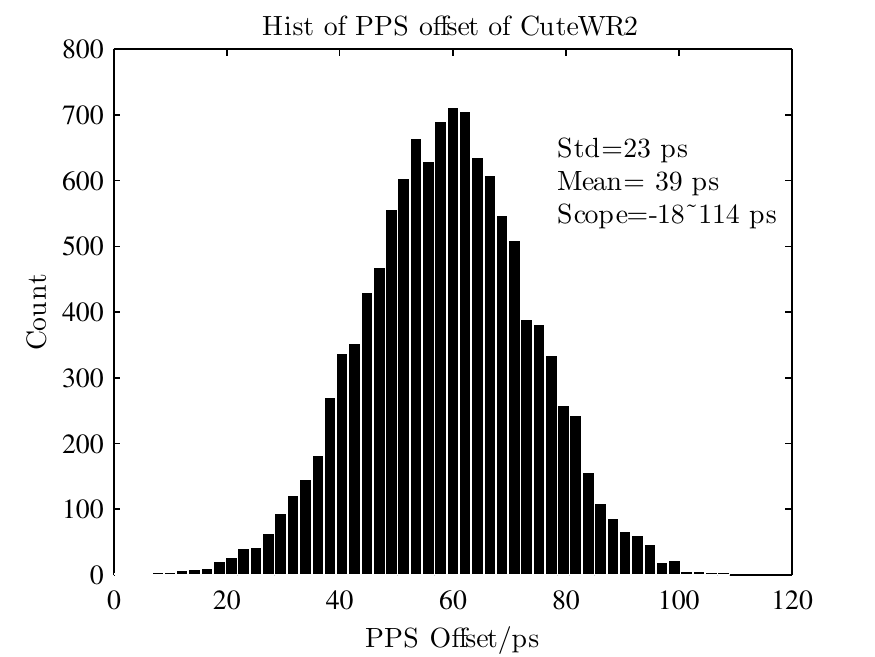}%
\label{fig:hist2}}
\hfill
\subfloat[PPS offset of CuteWR3]{\includegraphics[width=1.6in]{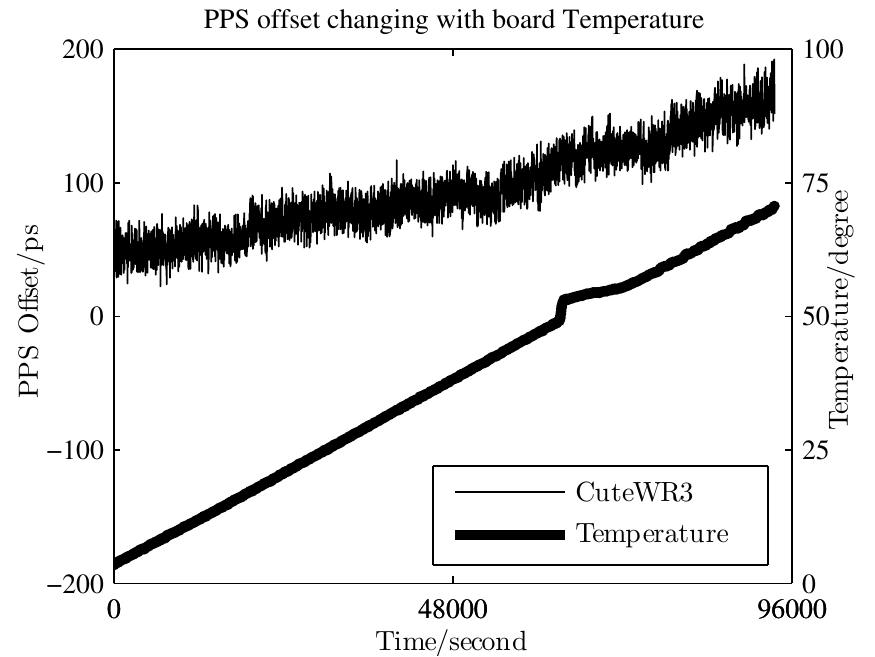}%
\label{fig:performance3}}
\subfloat[Histogram of CuteWR3]{\includegraphics[width=1.6in]{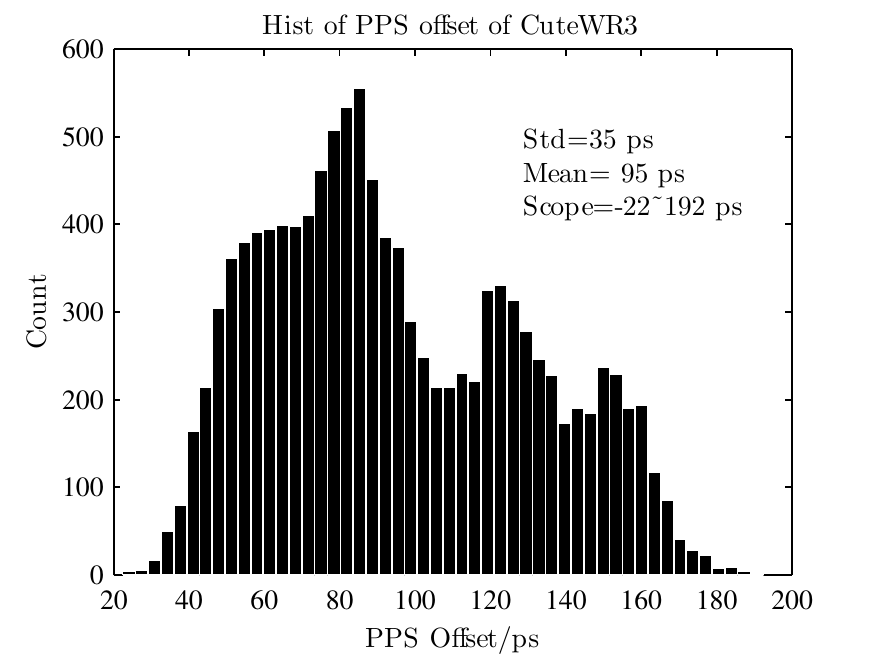}%
\label{fig:hist3}}
\caption{PPS offset between the master and the CUTE-WR slave nodes changing over time when CUTE-WR board temperature varying after fixed delays variation corrected and their histogram (fibers of 2km used).}
\label{fig:proformancetest}
\end{figure}

As seen in Figure \ref{fig:performance1}$\sim$\ref{fig:hist3}, the variations of the PPS offset have been reduced. Though the PPS offset of CuteWR3 is not completely corrected because of the components diversity, the variation is less than 150 ps and the standard deviation is smaller than 50 ps under a varying temperature between -10 and 55 $^{\circ}$C.

\section{Discussion}
Experiments show that fixed delays and $\alpha$ have dependency with temperature, that reduces the synchronization precision at different levels in WR network. The variation of fixed delays has a significant effect on the synchronization precision, which however can be corrected with a linear model in real time. A node by node calibration is needed to overcome components diversity if extreme precision is required.

The influence of the variation of $\alpha$ depends on the link length and the type of SFP modules. For short link, the effect is neglectable. However, it must be taken into consideration when long fibers are used. But the effect of the link temperature cannot be obtained in practice, making it impossible to be corrected. Two fibers with the same Tx wavelength can be an optional strategy. The influence of the spectrum temperature dependency of SFP laser diobe can be corrected using the same way as the fixed delays if needed. 
\section{Conclusion}
White Rabbit (WR) is designed to compensate temperature variation of the fibers. It has been tested and proven that such a compensation scheme works well in sub-ns precision. However the synchronization precision of WR under different temperature is influenced by the variation of $\alpha$ and fixed delays. Some measurements have been done and the results show that the variation of fixed delays may cause a 10 ps/$^{\circ}$C deviation, which can be easily to correct using a linear model and reduced to less than 2 ps/$^{\circ}$C for different CuteWR nodes, and the variation of $\alpha$ may also lead to a deviation less than 1 ps/($^{\circ}$C$\cdot$km), which can be ignored in the application of narrow temperature variation or short distance. The results also show that the precision\cite{cutewr:accuracy} of synchronization between WR switch and CUTE-WR nodes connecting with 2 kilometers fibers are smaller than 50 ps when CUTE-WR cards are under varying temperatures between -10 and 55 $^\circ$C.

\section*{Acknowledgment}
This work is supported by the National Science Foundation of China (No. 11275111). The authors would like to thank Huihai He, Javier Serrano, Maciej Lipinski, Erik Van Der Bij and the White Rabbit team at CERN for their help.

\bibliographystyle{IEEEtran}
\bibliography{IEEEabrv,./IEEEexample,./cutewr}


\end{document}